\documentclass{PoS}
\usepackage{amsmath}
\newcommand{\lms}{\Lambda_{\overline{\text{MS}}}}

\title{The asymmetry of the dimension two condensate}

\ShortTitle{The asymmetry of the dimension two condensate}

\author{\speaker{David Vercauteren}$^1$\\
        $^1$Department of Physics and Astronomy, Ghent University\\
        E-mail: \email{David.Vercauteren@UGent.be}}
\author{David Dudal$^1$\\
        E-mail: \email{David.Dudal@UGent.be}}
\author{John Gracey$^2$\\
        $^2$Theoretical Physics Division, Department of Mathematical Sciences, University of Liverpool\\
        E-mail: \email{gracey@liv.ac.uk}}
\author{Nele Vandersickel$^1$\\
        E-mail: \email{Nele.Vandersickel@UGent.be}}
\author{Henri Verschelde$^1$\\
        E-mail: \email{Henri.Verschelde@UGent.be}}

\abstract{I present recent analytical work concerning the electric-magnetic asymmetry in the dimension two condensate $\langle A_\mu^2\rangle$ in pure Yang--Mills theory. We reproduce qualitatively the lattice results previously found by Chernodub and Ilgenfritz.}

\FullConference{Light Cone 2010 - LC2010\\
		June 14-18, 2010\\
		Valencia, Spain}

\begin{document}

\section{Introduction}
Recent years have witnessed a great deal of interest
in the possible existence of mass dimension
two condensates in gauge theories, see for example
\cite{Gubarev:2000eu,Gubarev:2000nz,Verschelde:2001ia,Dudal:2002pq,Dudal:2003vv,Vercauteren:2007gx,Dudal:2005na,Boucaud:2001st,Furui:2005he,Gubarev:2005it,Browne:2003uv,Andreev:2006vy,RuizArriola:2006gq,Chernodub:2008kf} and references therein for approaches
based on phenomenology, operator product expansion,
lattice simulations, an effective potential
and the string perspective. There is special
interest in the operator
\begin{equation} \label{mingaugeorbit}
A_{\min}^2 = \min_{U\in SU(N)} \mathcal V^{-1} \int d^4x (A_\mu^U)^2 \;,
\end{equation}
which is gauge invariant due to the minimization
along the gauge orbit. It should be mentioned
that obtaining the global minimum is delicate
due to the problem of gauge (Gribov) ambiguities
\cite{Gribov:1977wm}. As is well known, local gauge
invariant dimension two operators do not exist in
Yang-Mills gauge theories. The nonlocality of \eqref{mingaugeorbit}
is best seen when it is expressed as \cite{Lavelle:1995ty}\footnote{We will always work in Euclidean spacetime.}
\begin{equation} \label{akwadraatlang}
A_{\min}^{2} =\int d^{4}x\left[ A_{\mu }^{a}\left( \delta _{\mu \nu }-\frac{\partial _{\mu }\partial _{\nu }}{\partial ^{2}}\right) A_{\nu}^{a} - gf^{abc}\left( \frac{\partial _{\nu}}{\partial ^{2}}\partial A^{a}\right) \left( \frac{1}{\partial^{2}}\partial {A}^{b}\right) A_{\nu }^{c}\right] + \mathcal O(A^{4}) \;.
\end{equation}
The relevance of the condensate $\langle A_\mu^2\rangle_\mathrm{min}$ was discussed
in \cite{Gubarev:2000eu,Gubarev:2000nz}, where it was shown that it can
serve as a measure for the monopole condensation
in the case of compact QED.

All efforts so far have concentrated on the Landau
gauge $\partial_\mu A_\mu = 0$. The preference for this particular
gauge fixing is obvious since the nonlocal expression
\eqref{akwadraatlang} reduces to the local operator $A^2_\text{min} = A_\mu^2$. In the case of a local operator, the Operator
Product Expansion (OPE) becomes applicable,
and consequently a measurement of the soft (infrared)
part $\langle A_\mu^2\rangle_\text{OPE}$ becomes possible. Such an
approach was followed in e.g. \cite{Boucaud:2001st} by analyzing
the appearance of $1/q^2$ power corrections in (gauge
variant) quantities like the gluon propagator or
strong coupling constant, defined in a particular
way, from lattice simulations. Let us mention
that already three decades ago attention was paid
to the condensate $\langle A_\mu^2\rangle$ in the OPE context \cite{Lavelle:1988eg}.

\begin{figure}\begin{center}
\includegraphics[width=6cm]{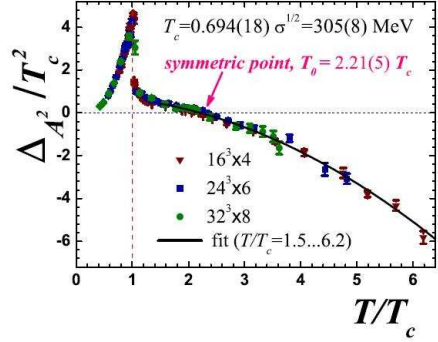}
\caption{Lattice results for the electric-magnetic assymetry $\Delta_{A^2}$ as function of the temperature for SU(2) pure gauge theory, as found in \cite{Chernodub:2008kf}. \label{maximplot}}
\end{center}\end{figure}

Recently, Chernodub and Ilgenfritz \cite{Chernodub:2008kf} have considered the asymmetry in the dimension two condensate. They performed lattice simulations, computing the expectation value of the electric-magnetic asymmetry in Landau gauge, which they defined as
\begin{equation}\label{ci}
\Delta_{A^2} = \langle g^2 A_0^2 \rangle - \frac1{d-1} \sum_{i=1}^{d-1} \langle g^2 A_i^2 \rangle\, .
\end{equation}
At zero temperature, this quantity must, of course, be zero due to Lorentz invariance\footnote{We shall deliberately use the term Lorentz invariance, though we shall be working in Euclidean space throughout this paper.}. Necessarily it cannot diverge as divergences at finite $T$ are the same as for $T=0$, hence this asymmetry is, in principle, finite, and it can be computed without renormalization, for all temperatures. Their results are depicted in \figurename\ \ref{maximplot}. At high temperatures, general thermodynamic arguments predict a polynomial behavior $\propto T^2$, and this is also what the authors of \cite{Chernodub:2008kf} found\footnote{A perturbative computation gives a positive proportionality constant, in contrary to what is erronously \cite{maxim} found in \cite{Chernodub:2008kf}. The lattice computations for $T<6\;T_c$ find a negative proportionality constant, so one would expect the real high-temperature behavior to start yet later.}. For the low-temperature behavior, however, one would expect an exponential fall-off with the lowest glueball mass in the exponent, $\Delta\sim e^{-m_{\text{gl}}T}$. Instead, they found an exponential with a mass $m$ significantly smaller than $m_\text{gl}$.

\section{$\langle A_\mu^2\rangle$ and $\Delta_{A^2}$ in the LCO formalism}
In order to get more insight in the behavior of the asymmetry, we have investigated it using the formalism presented in \cite{Verschelde:2001ia}. A meaningful effective potential for the condensation of the \emph{Local Composite Operator} (LCO) $A_\mu^2$
was constructed by means of the LCO method.
This is a nontrivial task due to the compositeness
of the considered operator. We consider pure
Euclidean SU(N) Yang--Mills theories with action
\begin{equation}
S_\text{YM+gf} = \int d^4x \left(\frac14 (F_{\mu\nu}^a)^2 + S_\text{gf} + b^a\partial_\mu A_\mu^a + \bar c^a\partial_\mu\mathcal D_\mu^{ab}c^b\right) \;.
\end{equation}
We couple the operator $A_\mu^2$ to the Yang--Mills action
by means of a source $J$:
\begin{equation}
S_J = S_\text{YM} + \int d^4x \left(\frac12J(A_\mu^a)^2 - \frac12\zeta J^2\right) \;.
\end{equation}
The last term, quadratic in the source $J$, is necessary
to kill the divergences in vacuum correlators like $\langle A^2(x)A^2(y)\rangle$ for $x\to y$, or equivalently in
the generating functional $W[J]$, defined as
\begin{equation}
e^{-W[J]} = \int [\text{fields}] e^{-S_J} \;.
\end{equation}
The presence of the LCO parameter $\zeta$ ensures a
homogenous renormalization group equation for
$W[J]$. Its arbitrariness can be overcome by making
it a function $\zeta(g^2)$ of the strong coupling constant
$g^2$, allowing one to fix $\zeta(g^2)$ order by order
in perturbation theory in accordance with the
renormalization group equation.

In order to access the electric-magnetic asymmetry, a second source $K_{\mu\nu}$ is coupled to the traceless part of $A_\mu^a A_\nu^a$. This second operator will not mix with $A_\mu^2$ itself, which allows control over the renormalization group of these two operators. Again a term quadratic in the new source must be added, introducing a second parameter $\omega(g^2)$ which can, again, be fixed order by order in accordance with the renormalization group equation. We have proven the all-order perturbative renormalizability of this extention of the formalism using the algebraic method based on the Ward identies \cite{Dudal:2009tq}.

In order to recover an energy interpretation,
the terms $\propto J^2$ and $k^2$ can be removed by employing a
Hubbard--Stratonovich transformation, amounting to inserting the following unities into the path integral:
\begin{equation} \label{unities}
1 = \int [d \sigma] e^{-\frac{1}{2\zeta} \int d^d x \left( \frac{\sigma}{g} + \frac{1}{2} A_\mu^2 - \zeta J \right)^2 } = \int [d \varphi_{\mu\nu}] e^{- \frac{1}{2\omega} \int d^d x \left( \frac{1}{g} \varphi + \frac{1}{2} A_\mu A_\nu -\omega\ k_{\mu\nu} \right)^2}\;,
\end{equation}
with $\varphi_{\mu\nu}$ a traceless field, leading to the action
\begin{equation}
S = S_\text{YM} + \int d^dx\left[\frac{1}{2\zeta} \frac{\sigma^2}{g^2} +  \frac{1}{2\zeta g}\sigma A_\mu^2 + \frac{1}{8\zeta} (A_\mu^2)^2 + \frac{1}{2\omega}\frac{\varphi_{\mu\nu}^2}{g^2} +  \frac{1}{2\omega g}\varphi_{\mu\nu} A_\mu A_\nu + \frac{1}{8\omega} (A^a_\mu A^a_\nu)^2\right] \;.
\end{equation}
Starting from this, it is possible to compute the effective potential $V(\sigma,\varphi_{\mu\nu})$, given the correspondences
\begin{equation}
\langle\sigma\rangle = -\frac g2\langle A_\mu^2\rangle \;, \quad \langle\varphi_{\mu\nu}\rangle = -\frac g2 \left\langle A_\mu A_\nu-\frac{\delta_{\mu\nu}}d A_\lambda^2\right\rangle \;.
\end{equation}

Now we determine the values of $\zeta$ and $\omega$ from the renormalization group equations for the sources $J$ and $K_{\mu\nu}$. For this, some anomalous dimensions and renormalization factors have to be computed up to one loop order higher than the intended loop order we are interested in. We have done this using the {\sc Mincer} algorithm. The final result is up to one-loop order:
\begin{equation}
\zeta = \dfrac{N^2 -1}{16 \pi^2} \left[\dfrac{9}{13} \dfrac{16 \pi^2}{ g^2 N} + \dfrac{161}{52} \right] \;, \qquad
\omega = \dfrac{N^2 -1}{16 \pi^2} \left[\dfrac{1}{4} \dfrac{16 \pi^2}{ g^2 N} + \dfrac{73}{1044} \right] \;.
\end{equation}

\section{Computation and minimalization of the action}
The effective potential $V(\sigma,\varphi_{\mu\nu})$ can now be computed using standard techniques. We have taken the background fields $\sigma$ and $\varphi_{\mu\nu}$ to have space-time independent vacuum expectation values and $\varphi_{\mu\nu}$ to be the traceless diagonal matrix $\operatorname{diag}(A,-\frac1{d-1}A,\ldots,-\frac1{d-1}A)$ \cite{Vercauteren:2010rk}.

\subsection{Low temperatures}
\begin{figure}\begin{center}
\includegraphics[width=5cm]{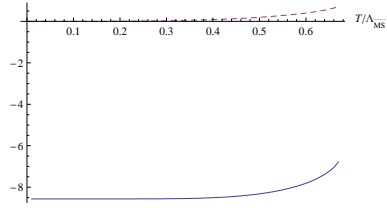}
\caption{The $\langle g^2A_\mu^2\rangle$ condensate (full line) and the assymetry $\Delta_{A^2}$ (dashed line) as functions of the temperature, in units $\lms$. \label{eerstefiguur}}
\end{center}\end{figure}

Computing the effective action up to one-loop order at zero temperature, yields only the minimum found in \cite{Verschelde:2001ia}, which is what we expect. For finite but still not too high temperatures, the potential can be minimized numerically. The result is depicted in \figurename\ \ref{eerstefiguur}. We see that the asymmetry rises at low temperatures, which agrees qualitatively with the findings of \cite{Chernodub:2008kf}. The low-temperature expansion of $\Delta_{A^2}$ reads
\begin{equation}
\Delta_{A^2} = (N^2-1) \frac{g^2\pi^2}{30} \left(1-\frac{85}{1044}\frac{g^2N}{(4\pi)^2}\right) \frac{T^4}{m^2} \;,
\end{equation}
and there is no correction to $\langle A_\mu^2\rangle$ at this order. Remark that we find a polynomial behavior $\propto T^4/m^2$ instead of an exponential. This does not agree with the lattice results, but in \cite{Chernodub:2008kf} the lowest temperatures reached were $T = 0.4\;T_c$, where our expansion is not valid anymore. This also does not agree with the thermodynamic argument that, in a theory with a massgap, one would expect exponential scaling. However, the asymmetry does not get a gauge invariant meaning by going to the Landau gauge.

\subsection{High temperatures}
At temperatures higher than $0.67\;\lms$, the minimum disappears. This signals a phase transition to the perturbative vacuum. In order to access this regime, it is possible to expand the effective potential for high temperatures, which yields
\begin{equation}
\langle g^2A_\mu^2 \rangle = g^2(N^2-1) \frac{T^2}4 \ , \qquad \Delta_{A^2} = g^2(N^2-1) \frac{T^2}{12} \;.
\end{equation}
This coincides with the perturbative result.

\begin{figure}\begin{center}
\includegraphics[width=4cm]{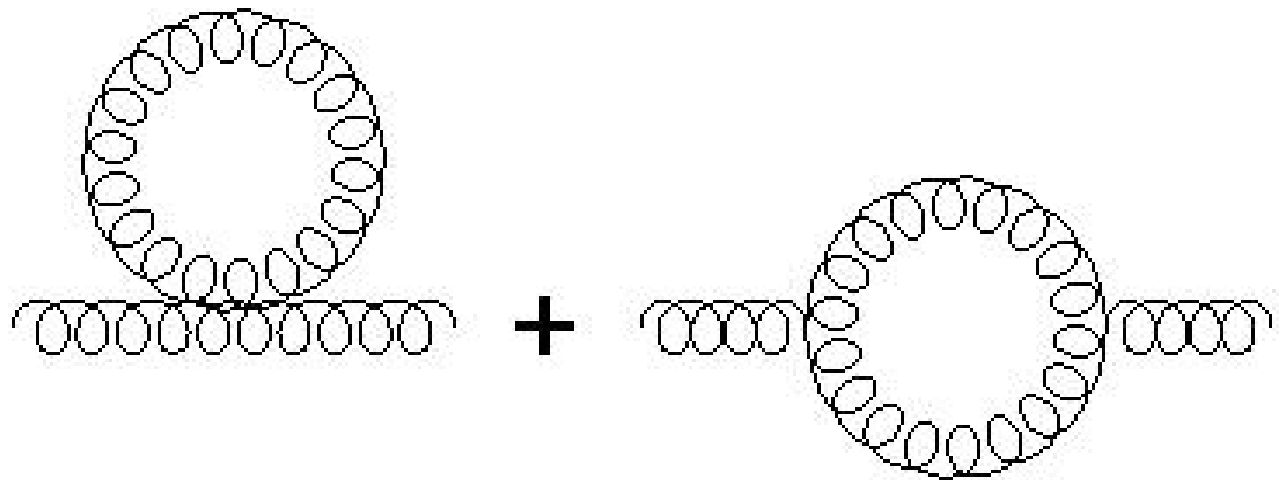} \hspace{3cm} \includegraphics[width=4cm]{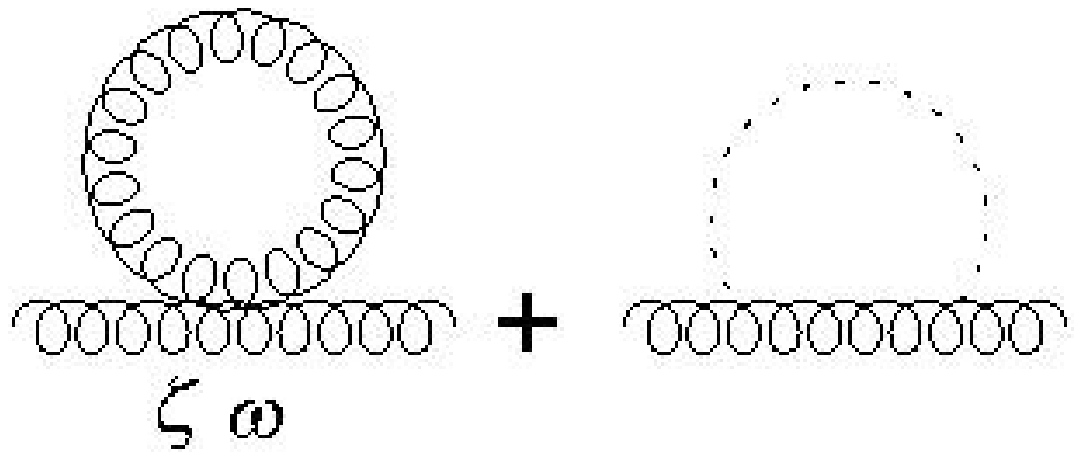}
\caption{\textbf{Left:} The diagrams giving the Debye mass in HTL resummation. The ghost loop is not necessary \cite{braatenpisarski}. \textbf{Right:} More diagrams that need to be resummed, coming from the $\sigma$ and $\varphi_{\mu\nu}$ parts of the LCO Lagrangian. The dotted line is the $\sigma$ or $\phi_{\mu\nu}$ propagator, and the labeled vertex depicts the additional vertices introduced by the formalism. \label{debye}}
\end{center}\end{figure}

In order to compute higher-order corrections to this, it is necessary to perform a Hard Thermal Loop (HTL) resummation \cite{andersenstrickland}, as nonresummed perturbation theory leads to a tachyonic mass in our case\footnote{The condensate and the mass generated by it differ by a negative multiplicative constant.}. In ordinary pure Yang--Mills theory at this order, HTL amounts to giving the timelike gluon a Debye mass $m_D^2 = \tfrac N3g^2T^2$, which effectively resums the hard (high momentum) contributions of the diagrams left in \figurename\ \ref{debye}. In our formalism, however, there are four additional vertices giving rise to four extra diagrams that need to be resummed, shown at the right in \figurename\ \ref{debye}. Computing these additional diagrams, it turns out that they exactly cancel the lowest-order contribution from the condensate. Solving in the large-$T$ limit yields
\begin{equation}
\langle g^2A_\mu^2\rangle = g^2(N^2-1)\left(\frac{T^2}4 - \frac{m_DT}{4\pi} + \cdots\right) \ , \qquad \Delta_{A^2} = g^2(N^2-1) \left(\frac{T^2}{12} - \frac{m_DT}{36\pi} + \cdots \right) \;,
\end{equation}
which is exactly what one would expect from perturbation theory.

\section{Conclusions}
The temperature dependence of the nonperturbative
dimension two condensate and its
electric-magnetic asymmetry have been studied
analytically. At low temperatures, we find
qualitative agreement with the lattice results.
At high temperatures, the perturbative vacuum
is recovered.

We wish to thank M.~N.~Chernodub for encouraging this
research.

\end{document}